\documentclass[12pt]{article}
\usepackage{amssymb}
\usepackage{amsfonts}
\usepackage{amsmath,amsthm}
\usepackage{graphicx,epsfig, color }
\usepackage{subfigure}

\oddsidemargin 10mm \numberwithin{equation}{section}
\def\be{\begin{equation}}
\def\ee{\end{equation}}
\begin{document}
\begin{center} {{\bf {Holographic thermalization in AdS-Gauss-Bonnet gravity for small entangled regions}}\\
 \vskip 0.50 cm
  {{ Hossein Ghaffarnejad \footnote{E-mail: hghafarnejad@semnan.ac.ir
 } }{ Emad Yaraie \footnote{E-mail: eyaraie@semnan.ac.ir
 } }{ Mohammad Farsam \footnote{E-mail: mhdfarsam@semnan.ac.ir
 } }}\vskip 0.2 cm \textit{Faculty of Physics, Semnan
University, P.C. 35131-19111, Semnan, Iran }}
\end{center}
\begin{abstract}
In this paper we study the propagation of entanglement entropy
after a global instantaneous quench on the CFT boundary of AdS
bulk. We consider the Gauss-Bonnet model as a higher curvature
gravity model for which we correct the RT(HRT) proposal to compute
the holographic entanglement entropy(HEE). To obtain an analytical
solution we perform an approximation approach which bounds our
computations to the small subregions and we compare its
thermalization regimes to the result of large subsystem case.
 We can see tsunami picture where
the evolution of entanglement breaks down for the large systems
 and so its details depends just on the shape and size of entangled region and also the used gravity model. We can
see the phase transition in this regime is always continuous
regardless the shape and size, in contrary with large subregions.
\end{abstract}

\section{Introduction}
The gauge/gravity duality or the AdS/CFT correspondence is a
conjecture that relates a gravity theory into the bulk to a
non-gravity theory
 (quantum field theory) on its boundary [1,2,3]. In fact this duality relates quantum physics of strongly coupled many-body systems to the classical
 dynamics of a gravity model which lives in one higher dimension. This makes the AdS/CFT a suitable way for the understanding and studying quantum
 gravity. Although, some recent works tried to generalize this conjecture to non-AdS gravity theories for which dual field theories must be invariant
  under other special scaling. For example in [4] the gravity model is used as a toy model in the condensed matter physics.
  The prominent work on this correspondence is RT proposal [5] to
study holographic entanglement entropy (HEE) which comes from the
similarity between the entanglement entropy as a non-local
observer in the boundary and black hole entropy in the bulk. In
this prescription entanglement entropy on the
 field theory side corresponds to a minimal surface in the bulk anchored to the entangled region, or in the other hand by geometrizing non-local
  observable on the filed side relate it to the bulk geometry. In fact observables on the field theory in this prescription can construct the bulk
  geometry. In a dynamical prescription of RT (or HRT [6]), minimal surface calculates over time until it equilibrates. In this situation the minimal
   surface can penetrate behind the event horizon and give us information from the black hole construction, however it can not still approach the
    singularity.
There are some limitations on the RT and HRT prescriptions which faces the reconstruction of the bulk to difficulty [7]. These limitations implies
 that HEE might not be a suitable probe to explore and construct the bulk geometry. There are two solutions to encounter this problem: using other
  useful probes such as mutual information [8], casual holographic information [9] and etc; the main problem about them is ambiguity about their dual
  field theory. Another approach is choosing an alternative gravity model such as higher derivative theories for the bulk instead of Einstein gravity
  model. The main problem arises here is that we can not use RT (or HRT) prescription anymore and we have to correct them. To correct this proposal we
  must consider some previous fundamental conditions like strong subadditivity [10]. It must be noted that evaluating new functional in higher curvature
  gravity model on the event horizon leads to the correction of black hole entropy in these theories. \\

In our work we pick up the second approach and by considering a higher derivative model of gravity (AdS-Gauss-Bonnet model) tries to explore
 the bulk. In this model it can be found that HEE extremal surface behaves like the Einstein gravity case and penetrates the horizon but doesn't
 reach to the singularity, however explores the bulk less than it. The extension of the RT proposal to the Gauss-Bonnet gravity has done in [11]
 and [12], and it generalized in [13] and [14] to any higher order theories and their dual field theories.
But dynamical process in the bulk during which entanglement entropy evolves is described by a global quantum quench on the boundary. Actually
 the initial static system in the pure AdS state perturbed by a time-dependent disturbance which is produced by injection of an uniform energy
 density at the initial time. These global quenches on the boundary are modeled by a null shell of matter which begins to collapse from the boundary
 at the initial time until formed to a black hole at the center of AdS spacetime [15,16,17]. If this injection is immediate then disturbance would be
 sharp at the moment corresponds to the collapsing of a very thin shell, while smooth perturbation is equivalent to a null shell with specified
 thickness.
Entanglement evolution from global quantum quench is studied in [18] for a large subsystem in 1+1-dimensional CFT. The result is that the
 entanglement entropy in large subsystems grows linearly in time and it is proportional to final thermal entropy density at saturation time.
 Saturation time at which the system equilibrated and entanglement entropy does't evolve anymore is proportional to the size of entangled region.
  This is clear enough about EPR pairs of entangled quasi-particles emitted from an initial state [18], but it must be more careful in many-body
  systems with strong interactions between the pairs. The AdS/CFT correspondence is a fundamental geometric approach to face with this subject.
  In [5] the entanglement entropy growth is obtained for large subsystem $A$ on the CFT boundary (with area $A_{\Sigma}$ on the boundary) by this
  correspondence as follows:
\begin{equation}
\Delta S_A(t)=v_Es_{eq}A_{\Sigma}t,~~~~t_{loc}\ll t\ll t_{sat},
\end{equation}
which denotes a linear and universal behavior in a time interval
between thermal equilibrium timescale and saturation time. At
saturation time
 ($t_{sat}$) collapsing null shell grazes the deepest point on the extremal surface into the bulk called "turning point" and after that extremal
 surface stop changing. Turning point ($z_t$) is proportional to the entangled region size on the boundary and so $t_{sat}=z_t\sim\ell$, as it noted
 in [18]. Another timescale, $t_{loc}$, represents time at which the system equilibrates thermally and after that no thermodynamic entropy adds
  to it, anymore. This timescale has an inverse relationship with final temperature of the system which corresponds to the characteristic wavelength
  of the thermal excitations, $\lambda_{th}$. This characteristic wavelength is proportional to the event horizon of the black hole, $z_h$, which is
  formed into the bulk. So as a conclusion $t_{loc}\sim1/T_{final}\sim\lambda_{th}\sim z_h$. In large subsystems since the size of region is much
   larger than wavelength of the thermal excitations $\ell\gg\lambda_{th}$, so these excitations have own effect on entanglement growth
   until the system equilibrate thermally.\\

$v_E$ given in (1.1), is depended  to dimension of spacetime $d$
such as follows [19,20].
\begin{equation}
v_E=\sqrt{\frac{d}{d-2}}\bigg[\frac{d-2}{2(d-1)}\bigg]^{\frac{d-1}{d}}.
\end{equation}
It is easy to find that for any dimension $v_E\leqslant1$, which the equality case happens for $d=2$. Due to this bound and the general form of (1.1)
 we can attribute $v_E$ as the speed of entanglement propagation. Of course $v_E$ is not a real physical velocity and $v_E\leqslant1$ is just a
 similarity to causality. In many recent works this is showed that for large subsystem this bounding for $v_E$ [21,22] is true. These similarities
  make it sense that we call $v_E$ the speed of entanglement growth which can be bounded by the velocity of light. As we mentioned above for $d=2$
  this velocity equals to 1 which means that entanglement entropy is propagated by a free streaming of particles moving at the speed of light regardless
  to possible interactions.
Attributing $v_E$ to the speed of entanglement growth leads to a creative picture, called "tsunami picture", which ascribed it to the propagation
of a wavefront on the CFT side [19,20]. In this picture the evolution of entanglement described on the boundary and by turning on the quantum quench
a wavefront starts propagating from the boundary of entangled region inward. This propagation continues until it covers the whole region at
the saturation point. In this case we can define the instantaneous rate of propagation as a dimensionless quantity:
\begin{equation}
\Re(t)=\frac{1}{s_{eq}A_{\Sigma}}\frac{dS_A}{dt}.
\end{equation}

By attention to tsunami picture this instantaneous rate is bounded by tsunami velocity like the velocity introduced in (1.2), or in the
 other word $\Re(t)\leqslant v_E$. From $v_E\leqslant1$ we conclude that $\Re_{max}\leqslant1$.
But the situation is different in small subsystems in which due to $t_{sat}\ll t_{loc}$ we have $\ell\ll\lambda_{th}$ that represents the
 size of entangled region is much smaller than thermal excitations and therefore interactions have not any effect on entanglement evolution.
 In the other words, the system becomes saturated long before the thermal excitations could be effective. Thus, entropy evolution only happens in
 the initial time interval for large system and doesn't proceed more, so tsunami picture (1.1) breaks down here. Therefore, $\Re(t)$ is not bounded
 to tsunami velocity and can exceed the speed of light. This new situation makes the entanglement growth dependent on the details of gravity model
 and type of the quench, instead of interactions and thermal excitations. This again justifies the using of various gravity model for the bulk.\\

Layout of this paper is as follows: In section 2 we introduced
Einstein-Hilbert gravity modified with AdS-Gauss-Bonnet higher
order derivative counterpart in 5-dimensional curved space times
and describe a spherically symmetric static metric solution of the
model which has black hole topology. At the beginning of section 3
we explain our approximation approach which is the main procedure
we perform here for small subsystems to obtain the leading
behavior of entanglement evolution. By this approach we obtain the
explicit forms of our quantities to a large extent, which
describes the evolution of system. Because of unnecessary
complicated computations we just perform our calculations for the
strip region. In section 4 we try to analyze various regimes of
thermalization in this gravity model for small subsystems and
compare them with the main thermalization regimes for large
subsystems and describe their similarities and differences. At
last in section 5 we have a conclusion of the results and also
present some outlooks and further works.

\section{Gauss-Bonnet model of gravity in AdS space}

Gauss-Bonnet gravity model is obtained from the second order equations of motion due to the higher order derivatives terms in the
 action, Indeed it is the simplest form of the Lovelock gravity model. By attention to the AdS/CFT correspondence, these higher order
 derivative terms appear as a quantum or stringy corrections in the classical action as the correction of the curvature. By adding
 Gauss-Bonnet term to the uncharged Einstein theory of gravity we lead to a five-dimension Lovelock model with the following action [11]:
\begin{equation}
S=\frac{1}{16\pi G_N^{(5)}}\int d^5x\sqrt{-g}\bigg(R-2\Lambda+\frac{\lambda_{GB}L^2}{2}\mathcal{L}_{GB}\bigg)
\end{equation}
in which $\mathcal{L}_{(4)}$ is the Gauss-Bonnet term,
\begin{equation}
\mathcal{L}_{GB}=R_{\alpha\beta\gamma\delta}R^{\alpha\beta\gamma\delta}-4R_{\alpha\beta}R^{\alpha\beta}+R^2,
\end{equation}
and $G_N^{(5)}$ is the Newtonian constant in 5-dimension, $R$ is
Ricci scalar, $L$ is the radius of anti-de Sitter space which is
related to the cosmological constant via $\Lambda=-6/L^2$ in five
dimension, $\lambda_{GB}$ is the coupling constant in the
Gauss-Bonnet gravity which must satisfy the following constraint
due to the causality of dual field theory in $d$-dimension [23]:
\begin{equation}
-\frac{(3d-1)(d-3)}{2(d+1)^2}\leq\lambda_{GB}\leq\frac{(d-3)(d-4)(d^2-3d+8)}{2(d^2-5d+10)^2},
\end{equation}
where in our case for $d=5$ it reduces to
\begin{equation}
-0.39\lesssim\lambda_{GB}\leq0.18.
\end{equation}
The black hole solution for AdS-GB (Anti de Sitter Gauss-Bonnet
model) of (2.1) by attention to [24] is
\begin{equation}
ds^2=-f(r)dt^2+\frac{dr^2}{f(r)}+r^2\big[d\theta^2+sin^2\theta(d\phi^2+sin^2\theta d\psi^2)\big],
\end{equation}
for which two solutions could be acceptable mathematically:
\begin{equation}
f(r)=\frac{r^2}{2\lambda_{GB}}\bigg(1\pm\sqrt{-\frac{4\lambda_{GB}}{L^2}+\frac{32\lambda_{GB}M}{3\pi r^4}+1}\bigg)+1,
\end{equation}
where $M$ represents the mass of black hole. It must be noticed
that we will ignore the positive one because it contains ghost
solution and so it is unstable. By defining a new coordinate as
$z=L^2/r$, for which the singularity $r=0$ sits at infinity and
AdS boundary would be $z=0$, we can transform the frame to the
Eddington-Finkelstein coordinate (see [26] and references therein
) with
\begin{equation}
dt=d\upsilon+\sqrt{f_0}\frac{dz}{f(z)},
\end{equation}
where,
\begin{equation}
f(z)=\frac{1}{2\lambda_{GB}}\bigg[1-\sqrt{1-4\lambda_{GB}(1-Mz^4)}\bigg],
\end{equation}
with $f_0$ for near the boundary value of $f(z)$, namely
\begin{equation}
f_0=\lim_{z\rightarrow0}f(z)=\frac{1}{2\lambda_{GB}}\bigg(1-\sqrt{1-4\lambda_{GB}}\bigg),
\end{equation}
which by attention to (2.4) in $d=5$ restricted as $0.77\lesssim f_0\lesssim1.31$. By these considerations
we lead to a new form of our metric as follows:
\begin{equation}
ds^2=\frac{L^2}{z^2}\bigg(-\frac{f(z)}{f_0}~d\upsilon^2-\frac{2}{\sqrt{f_0}}~dz d\upsilon+d\vec{x}~^2\bigg),
\end{equation}
where $d\vec{x}^{~2}=d\theta^2+sin^2\theta(d\phi^2+sin^2\theta d\psi^2)$. Note that the location of the event horizon, which obtained
by $f(z_h)=0$, by attention to (2.8) would be $z_h=M^{-1/4}$. Also the Hawking temperature of this solution can be achieved as [25]:
\begin{equation}
T=\frac{M^{1/4}}{\pi L^2\sqrt{f_0}}.
\end{equation}
In a Vaidya form, $f(z)$ from (2.8) has a dependence on a new time
coordinate, $\upsilon$, via the mass function and would be changed
like (see for instance [26, 27, 28]):
\begin{equation}
f(z,\upsilon)=\frac{1}{2\lambda_{GB}}\bigg[1-\sqrt{1-4\lambda_{GB}(1-M(\upsilon)z^4)}\bigg].
\end{equation}
In this work we consider time-dependence of the mass as step function, $M(\upsilon)=M\theta(\upsilon)$. It represents a shock wave with zero
 thickness which suddenly collapses at $\upsilon=0$. Transformation from a static AdS space-time to a black hole solution is described by a quench
 process for which $\upsilon<0$ implies the pure solution and this process is carried by a thin null shell lies at $\upsilon=0$.
\section{Time dependant behavior of entanglement entropy}
According to the prescription of Ryu and Takayanagi [5] in the
context of AdS/CFT correspondence, entanglement entropy of a
region $A$ on the field theory boundary is computed by:
\begin{equation}
S_A=\frac{1}{4G_N^{(d)}}min[Area(\Gamma_A)],
\end{equation}
where $G_N^{(d)}$ is the Newtonian constant in the AdS spacetime
and $\Gamma_A$ is a $d$-dimensional surface in the bulk which is
bounded to the CFT boundary such that $\partial\Gamma_A=\partial
A$. In time dependant version of this proposal called HRT [6] the
condition of minimal area surface would be generalized by extremal
surface,
\begin{equation}
S_A=\frac{1}{4G_N^{(d)}}~ext[Area(\Gamma_A)].
\end{equation}
We can consider region $A$ as different geometric surfaces and then it would be possible studying entanglement entropy explicitly. For
simplicity reasons we just consider the case for a rectangular strip surface on the boundary. In a $(d-1)$-dimensional strip region on the boundary
 with width $\ell$ and length $\ell_{i\bot}\rightarrow\infty$ ($i=1...d-2$) the boundary coordinates defines by:
\begin{equation}
x\in\big[-\frac{\ell}{2},\frac{\ell}{2}~\big],~~~~x_i\in\big[-\frac{\ell_{i\bot}}{2},\frac{\ell_{i\bot}}{2}~\big],
\end{equation}
and we can write $d\vec{x}^{~2}=dx^2+dx_i^2$. It must be noticed that extremal surface in this case is invariant under translations in the direct
of $x_i$ and we can parameterize it with embedding functions $x(z)$ and $\upsilon(z)$ by below boundary conditions:
\begin{equation}
x(0)=\pm\frac{\ell}{2}~,~~~~\upsilon(0)=t,~~~x(z_t)=0,
\end{equation}
where $z_t$ is a specific value for holographic direction on the extremal surface called "turning point" for which ($z_t$,$\upsilon_t$) is
deepest point within the bulk and completely fixes the extremal surface.\\
Now, the above considerations help us to define the area of
extremal surface as follows
\begin{equation}
Area(\Gamma_A)\equiv \mathcal{A}(t)=\int_{\Sigma}d^{d-2}\zeta\sqrt{\gamma},
\end{equation}
in which $\zeta$ represents world-volume coordinate (all coordinates except coordinates which defined as embedding functions) and $\gamma$ is
determinant of induced metric on surface $\Sigma$. Equation of motions simply derived from the extremizing of the functional in (3.5), but in our
case for (2.12) we have to use an approximation technique to study leading order of the solution by considering small subsystems (when $z_t\ll z_h$)
 which makes possible to solve it analytically [29].\\
$\bullet$ \textbf{Perturbation method to solve non-linear equations:}\\
Consider a general form of the functional as $\mathcal{L}[\psi(z),\epsilon]$ where $\psi(z)$ denotes all embedding functions defined in
the functional and $\epsilon\ll1$ is a dimensionless parameter that plays the role of perturbation parameter. By this definition we can expand
 the functional and embedding functions with respect to $\epsilon$ as follows,
\begin{gather}
 \nonumber \mathcal{L}[\psi(z),\epsilon]=\mathcal{L}^{(0)}[\psi(z)]+\epsilon \mathcal{L}^{(1)}[\psi(z)]+\mathcal{O}(\epsilon^2),\\
\psi(z)=\psi^{(0)}(z)+\epsilon\psi^{(1)}(z)+\mathcal{O}(\epsilon^2).
\end{gather}
All embedding functions could be obtained by solving the equations of motion from the functional but since the equations are non-linear in general,
solving them are very difficult and even impossible. By writing the functional and embedding functions order by order like (3.6) and keeping only the
first order and neglecting terms with higher order (because the key observation is at first order), the expansion of the functional leads to
\begin{gather}
\nonumber \mathcal{A}[\psi(z)]=\int dz\mathcal{L}[\psi(z),\epsilon]=\int dz\mathcal{L}^{(0)}[\psi^{(0)}(z)]+\epsilon\int dz\mathcal{L}^{(1)}
[\psi^{(0)}(z)] \\
+\int dz~\epsilon\frac{\partial\mathcal{L}(\psi(z),\epsilon)}{\partial\epsilon}\bigg|_{\epsilon=0}.
\end{gather}
The last term in the above equation could be written in the following form for which we use of the Lagrangian variation method:
\begin{equation}
\frac{\partial\mathcal{L}(\psi(z),\epsilon)}{\partial\epsilon}\bigg|_{\epsilon=0}=\psi^{(1)}(z)\frac{\partial\mathcal{L}}{\partial\psi}
\bigg|_{\psi=\psi^{(0)}}=\psi^{(1)}(z)\bigg(\frac{\partial\mathcal{L}}{\partial\psi}-\frac{d}{dz}\frac{\partial\mathcal{L}}{\partial\psi^{\prime}}
\bigg)_{\psi^{(0)}}
\end{equation}
By considering the Euler-Lagrange equation of motion the above term vanishes and the on-shell result of (3.7) would be
\begin{equation}
\mathcal{A}_{\text{on-shell}}[\psi(z)]=\int dz\mathcal{L}^{(0)}[\psi^{(0)}(z)]+\epsilon\int dz\mathcal{L}^{(1)}[\psi^{(0)}(z)]+\mathcal{O}(\epsilon^2).
\end{equation}
The interesting point of this result is that the solving of first order expansion just needs the zeroth order of the embedding functions. \\
\section{The growth of entanglement entropy in AdS-Gauss-Bonnet gravity}
Since AdS-Gauss-Bonnet model is a higher derivative theory of
gravity so we can not use RT (3.1) or HRT (3.2) prescriptions to
evaluate entanglement entropy. In this situation we must modified
our proposal according to [11] and [12] by defining:
\begin{equation}
\mathcal{S}_{EE}=\frac{1}{4G_N^{(5)}}\int_{\Sigma}d^3\zeta\sqrt{\gamma}(1+\lambda_{GB}L^2R_{\Sigma})+\frac{1}{2G_N^{(5)}}
\int_{\partial\Sigma}d^2\zeta~\lambda_{GB}L^2\sqrt{h}K,
\end{equation}
in which the entanglement entropy for an arbitrary region on CFT side would be obtained by extremizing the above statement
for $\mathcal{S}_{EE}$. Also $\Sigma$ is a three dimensional surface into the bulk anchored to 2-dimensional boundary, $\zeta$ represents
 world-volume coordinate on the both terms (where $d^3\zeta=dzdx_1dx_2$ in the first integral and $d^2\zeta=dx_1dx_2$ in the second one),
  $\gamma$ is determinant of induced metric on the surface $\Sigma$ which constructs $R_{\Sigma}$ as the Ricci scalar for the intrinsic
  geometry on $\Sigma$. The second term called the Gibbons-Hawking-York boundary term (GHY term) [30] is added to provide a good variational
   principle in extremizing our functional. In this boundary term induced metric denotes by $h$ and $K$ is the trace of the extrinsic curvature
    of $\Sigma$.
For a rectangular strip on the boundary described in section 3 with translational invariance in directions $x_1$ and $x_2$ and with width $\ell$
in the other direction $x$ ($-\ell/2\leq x\leq\ell/2$), the extremal surface bound to this area would be parameterized by the bulk direction $z$,
 and therefore our embedding functions would be $x=x(z)$ and $\upsilon=\upsilon(z)$. By these considerations we can rewrite (2.10) as follows:
\begin{equation}
ds^2=\gamma_{ab}dx^adx^b=\frac{L^2}{z^2}\bigg[\bigg(x^{\prime2}-\frac{f(z)}{f_0}\upsilon^{\prime2}-\frac{2}{\sqrt{f_0}}\upsilon^{\prime}\bigg)
dz^2+dx_1^2+dx_2^2\bigg],
\end{equation}
which actually represents induced metric on the co-dimension surface and prime denotes derivative with respect to $z$, also the boundary conditions
 in (3.4) are the same here as well. This induced metric leads us to
\begin{equation}
\sqrt{\gamma}=\frac{L^3}{z^3\sqrt{f_0}}\sqrt{\mathcal{G}},
\end{equation}
where
\begin{equation}
\mathcal{G}=\mathcal{G}(z)\equiv f_0x^{\prime2}-f(z)\upsilon^{\prime^2}-2\sqrt{f_0}\upsilon^{\prime}.
\end{equation}
After calculating the Ricci scalar, $R_{\Sigma}$, which is constructed from the above induced metric we get
\begin{equation}
R_{\Sigma}=-\frac{2f_0}{\mathcal{G}^2L^2}(z\mathcal{G}^{\prime}+3\mathcal{G}).
\end{equation}
By attention to $R_{\Sigma}$, the second term in the first integral in (4.1) could be rewritten as the following form because of the future
 simplifying purposes:
\begin{equation}
\lambda_{GB}L^2\sqrt{\gamma}R_{\Sigma}=\frac{2\lambda_{GB}L^3\sqrt{f_0}}{z^3\sqrt{\mathcal{G}}}+\frac{d}{dz}\bigg(\frac{4\lambda_{GB}L^3
\sqrt{f_0}}{z^2\sqrt{\mathcal{G}}}\bigg).
\end{equation}
This simplified form shows its important role during the calculation of GHY term in (4.1) (see the details in appendix A). The result is as below:
\begin{equation}
\lambda_{GB}L^2\sqrt{h}K=-\frac{2\lambda_{GB}L^3\sqrt{f_0}}{z^2\sqrt{\mathcal{G}}}
\end{equation}
As we can see GHY term cancels exactly with the second term in (4.6) when all terms are placed in (4.1) and the final modified action  which
needs to be extremized to obtain the entanglement entropy would be:
\begin{equation}
\mathcal{S}_{EE}=\frac{L^3}{4G_N^{(5)}\sqrt{f_0}}\int_{\Sigma}dx_1dx_2\frac{dz}{z^3}\bigg(\sqrt{\mathcal{G}}+\frac{2\lambda_{GB}f_0}{\sqrt{\mathcal{G}}}
\bigg).
\end{equation}
As it mentioned before, deriving the equations of motion and reaching to an explicit form of the entanglement entropy needs to apply perturbative
technique. In the approximation of small subsystems, for which $z_t\ll z_h$, the embedding functions and the functional could be expanded with respect
to a perturbation parameter, similar to (3.6). By attention to the on-shell expansion (3.9) we just need zeroth order of the embedding functions to
 expand the functional to first order.
At first place to expand (2.12) we can apply $f(z_h)=0$ and obtain time-dependent mass function as $M(\upsilon)=M\theta(\upsilon)=\theta(\upsilon)/z_h^4$. By attention to (2.9) we take $2\lambda_{GB}f_0=\alpha$ corresponds to $f_0=2/(2-\alpha)$ which leads to the following bound by regarding (2.4):
\begin{equation}
-0.6\lesssim\alpha\lesssim0.47,
\end{equation}
and by considering $z_t\ll z_h$ we can expand (2.12) as follows:
\begin{equation}
f(z,\upsilon)=a+b\bigg(\frac{z}{z_t}\bigg)^4\theta(\upsilon)\epsilon+c\bigg(\frac{z}{z_t}\bigg)^8\theta(\upsilon)\epsilon^2+\mathcal{O}(\epsilon^3),
\end{equation}
in which $\epsilon\equiv(z_t/z_h)^4\ll1$ is perturbation parameter and
\begin{equation}
a=\frac{2}{2-\alpha},~~~b=-\frac{1}{1-\alpha},~~~c=\frac{\alpha(2-\alpha)}{4(1-\alpha)^3}.
\end{equation}
By attention to (4.9) none of coefficients in (4.11) doesn't
suffer from any singularity problem. If we ignore the second and
higher order perturbation terms and keep only the first order, we
can expand the action (4.8) as follows:
\begin{equation}
\mathcal{S}_{EE}=\frac{1}{4G_N^{(5)}}\int_{\Sigma}dz\bigg(\mathcal{L}^{(0)}+\epsilon\theta(\upsilon)\mathcal{L}^{(1)}\bigg),
\end{equation}
in which, by setting $L=1$,
\begin{equation}
\mathcal{L}^{(0)}=\frac{A_{\Sigma}}{z^3\sqrt{f_0}}\bigg(\sqrt{\mathcal{G}_0}+\frac{\alpha}{\sqrt{\mathcal{G}_0}}\bigg),
\end{equation}
and
\begin{equation}
\mathcal{L}^{(1)}=\frac{-A_{\Sigma}}{z^3\sqrt{f_0}}\frac{\upsilon^{\prime2}b}{2\sqrt{\mathcal{G}_0}}\bigg(1-\frac{\alpha}{\mathcal{G}_0}\bigg),
\end{equation}
where $A_{\Sigma}=\int_{\Sigma}dx_1dx_2=4\ell_1\ell_2$ ($\ell_1,\ell_2\rightarrow\infty$ are the lengthes of enough long strip in $x_1$ and $x_2$
directions) is a constant because the functional is independent from $x_1$ and $x_2$. Also $\mathcal{G}_0$ is the value of $\mathcal{G}$ for
$\epsilon=0$:
\begin{equation}
\mathcal{G}_0=f_0x^{\prime2}-2\sqrt{f_0}\upsilon^{\prime}-\upsilon^{\prime2}a.
\end{equation}
To apply perturbation approach we need the zeroth order of the embedding functions, so it must be necessary to expand them for small $\epsilon$
and ignore all terms except zeroth order terms. At first by using (2.7) and (4.10) we lead to
\begin{equation}
\upsilon(z)=t-\frac{1}{\sqrt{a}}z+\bigg(\frac{b}{a\sqrt{a}}\frac{\theta(\upsilon)}{5z_t^4}z^5\bigg)\epsilon+\mathcal{O}(\epsilon^2)~~~~
\text{for}~\upsilon>0.
\end{equation}
Since we need only the zeroth order so $\upsilon^{(0)}(z)=t-\frac{1}{\sqrt{a}}z$ and $\upsilon^{(0)\prime}=-\frac{1}{\sqrt{a}}$, for which prime
denotes derivative with respect to the holographic coordinate $z$. By plugging these results into $\mathcal{L}^{(0)}$ and by solving the Euler-Lagrange
equations we find:
\begin{equation}
x^{(0)\prime}=\frac{1}{\sqrt{a}}\frac{\pm\chi z}{\sqrt{a-\chi^2z^2}},
\end{equation}
since the evolution of $x(z)$ and $z$ are inversely related ($\frac{dx}{dz}<0$), so without loss of generality we can consider minus sign.
Also $\chi=\chi(z,\kappa)$ in which $\kappa$ is a constant of motion from the Euler-Lagrange equations. Actually since $\mathcal{L}^{(0)}$ doesn't
include $x$ explicitly, so the equation of motion has a constant we denoted by $\kappa$ and by using this constant $\chi$ would be the real root of a
 polynomial function whose explicit form is as below:
\begin{equation}
\chi=\frac{12^{1/3}}{6z}\bigg(\frac{\chi_0}{\alpha}+\frac{a(\alpha-1)12^{1/3}}{\chi_0}\bigg),
\end{equation}
for which,
\begin{equation}
\chi_0=\bigg[a\alpha^2\bigg(9\kappa z^3+\sqrt{\frac{81\kappa^2\alpha z^6-12a(\alpha-1)^3}{\alpha}}\bigg)\bigg]^{\frac{1}{3}}.
\end{equation}
Regardless the details of $\chi$, we can evaluate this function for $z=z_t$. In $z=z_t$ we have $x^{(0)\prime}\rightarrow\infty$ so
 $\chi(z_t)=\sqrt{a}/z_t$ and therefore constant of motion could be computed for any $\alpha$. So (4.17) could be evaluated as follows
\begin{equation}
x^{(0)\prime}=\frac{1}{\sqrt{a}}\frac{-(\frac{z}{z_t})}{\sqrt{1-(\frac{z}{z_t})^2}}.
\end{equation}
This result leads us to the zeroth order of another embedding function,
\begin{equation}
x^{(0)}(z)=\frac{\ell}{2}-\frac{z_t}{\sqrt{a}}\bigg(1-\sqrt{1-\big(\frac{z}{z_t}\big)^2}\bigg),
\end{equation}
which by reminding the condition $x(z_t)=0$ we find the relation between length of entangled region and turning point as
\begin{equation}
\ell=2z_t/\sqrt{a}.
\end{equation}
Since we are interested in the evolution of entanglement entropy
so we bound us to compute the second part of (4.12) as follows,
\begin{equation}
\Delta S(t)=S^{(1)}(t)=S(t)-S_{vac}=\frac{1}{4G_N^{(5)}}\int_{0}^{z_t}dz~\epsilon\theta(\upsilon)\mathcal{L}^{(1)}[x^{(0)}(z),\upsilon^{(0)}(z)],
\end{equation}
in which $\mathcal{L}^{(1)}$ is a function of the zeroth order of the embedding functions defined in (4.14). In the zeroth order the spacetime is
 static so all extremal surfaces lie on a constant-$t$ slice ($t(z)=t$) [29]. Solving (4.22) we change differential parameter to $\upsilon(> 0)$ for
 which $\theta(\upsilon)=1$ it turns to
\begin{equation}
S^{(1)}(t)=\frac{1}{4G_N^{(5)}}\int_{t-z_t}^{t}d\upsilon~\epsilon\mathcal{L}^{(1)}[x^{(0)}(z),\upsilon^{(0)}(z)].
\end{equation}
Now according to the limit of above integral, two situations may be arisen. Indeed, time dependant behavior of the entanglement entropy is
different before and after the specific time-scale appropriate to $z_t$ called "saturation time" at which the null shell grazes the turning
point; in the other words at the saturation time $\upsilon(z_t)=0$ which by putting that in $\upsilon^{(0)}(z)$ we get to $t_{sat}=z_t$. Since
from (4.22) the turning point is appropriate with the size of the entangled region, then the saturation time would be increased by increasing of
 the sub-region size. By this saturation time the following two cases could be studied:\\
$\textbf{a)}$ If $t<t_{sat}$ then the range of the integral varies
from zero to $t$ and the final result is a time function
\begin{gather}
\nonumber \Delta S(t<t_{sat})=\frac{(2-\alpha)^{3/2}A_{\Sigma}z_t^2}{48\sqrt{2}G_N^{(5)}z_h^4}\times\\ \bigg[1+\frac{2\alpha}{5(1-\alpha)}-
\bigg(1-\big(\frac{t}{z_t}\big)^2\bigg)^{3/2}\bigg(1+\frac{\alpha}{1-\alpha}\frac{3(\frac{t}{z_t})^2+2}{5}\bigg)\bigg],
\end{gather}
$\textbf{b)}$ If $t>t_{sat}$ then the evolution which is started
at $t=0$ stops after the saturation time. We can simply put
$t=t_{sat}$ in (4.25) and get to
\begin{equation}
\Delta S(t>t_{sat})=\frac{(2-\alpha)^{3/2}A_{\Sigma}z_t^2}{48\sqrt{2}G_N^{(5)}z_h^4}\bigg[1+\frac{2\alpha}{5(1-\alpha)}\bigg],
\end{equation}
in which we used the parameters defined in (4.11). As we know after the saturation time entanglement entropy doesn't change anymore we can
define $\Delta S(t>t_{sat})=\Delta S_{sat}$ and rewrite (4.25) as
\begin{equation}
\Delta S(t)=\Delta S_{sat}\bigg[1-\frac{\bigg(1-\big(\frac{t}{z_t}\big)^2\bigg)^{3/2}}{1+\frac{2\alpha}{5(1-\alpha)}}\bigg(1+\frac{\alpha}{1-\alpha}
\frac{3(\frac{t}{z_t})^2+2}{5}\bigg)\bigg].
\end{equation}
There is a dimensionless quantity as we introduced in the introduction which is useful to study the entanglement entropy evolution. This quantity
represents the instantaneous rate of this growth by factorizing the aspects of the system such as the size of the region (or total number of degrees
of freedom). A system with the bigger size has more degrees of freedom and so the faster speed of growth for entanglement entropy. This rate of
 entanglement growth is defined by [19,20],
\begin{equation}
\Re(t)=\frac{1}{s_{sat}A_{\Sigma}}\frac{d(\Delta S(t))}{dt},
\end{equation}
where $s_{sat}=\Delta S_{sat}(t)/V_A$ is the equilibrium entropy density of the system which happens after the saturation time and $V_A$ is the
 volume of the entangled region $A$. By using $A_{\Sigma}=4\ell_1\ell_2$ and $V_A=\ell\ell_1\ell_2$ and also $\ell=2z_t/\sqrt{a}=z_t\sqrt{2(2-\alpha)}$
 we reach to
\begin{equation}
\Re(t)=\frac{\ell}{4\Delta S_{sat}}\frac{d(\Delta S(t))}{dt}=\frac{3t\sqrt{1-\big(\frac{t}{z_t}\big)^2}}{4z_t\big(1+\frac{2\alpha}{5(1-\alpha)}\big)}
\bigg(1+\frac{\alpha}{1-\alpha}\big(\frac{t}{z_t}\big)^2\bigg)\sqrt{2(2-\alpha)}.
\end{equation}
For studying the entanglement entropy growth schematically, we plotted these analytic results for the various values of the Gauss-Bonnet coupling
constant from (2.4) in figure (1.a). Actually time-dependant function of entanglement entropy in the first order, $\Delta S(t)=S^{(1)}(t)$, behaves
 like (4.27) for $\alpha$ which can be varied like (4.9). The case for $\alpha=0$ corresponds to $\lambda_{GB}=0$ for which the effect of higher order
  Gauss-Bonnet vanishes plotted by the gray line. As a result, by increasing coupling constant $\lambda_{GB}$ for a fixed size of entangled region,
  saturation value of entanglement entropy decreased at a same saturation time.
In figure (1.b) we plotted the diagram of $\Re(t)$ for the same values of $\alpha$ we choose in (1.a), we can see that although they all never cross
the speed of light, but the shape of instantaneous rate function $\Re(t)$ is changed slightly by altering the coupling constant of our gravity model.
As we can see by increasing $\alpha$ the maximum value of $\Re(t)$ decreases until around $\alpha=0.4$ (or $\lambda_{GB}=0.16$) after which it
starts increasing again. In the other words, $\Re_{max}$ is a function of $\lambda_{GB}$ which has a minimum for $\lambda_{GB}=0.16$. Table 1
shows the changing of $\Re_{max}$ for (4.9) and by considerations mentioned in figure 1.
\begin{table}[h!]
  \begin{center}
    \label{tab:table1}
    \begin{tabular}{l|c|c} 
      $\mathbf{\alpha}$ & \textbf{$t_{max}$} &\textbf{$\Re_{max}$} \\
      \hline\hline
      -0.6 & 0.06283518555& 0.8379076380 \\
      -0.4 & 0.06487247509& 0.8057683980 \\
      -0.2 & 0.06747391448& 0.7757202720 \\
      0 & 0.07071067812& 0.7499999998 \\
      0.2 & 0.07444781001& 0.7321081323 \\
      0.4 &0.07825422900 & 0.7267219642 \\
      0.41 &0.07843687254 & 0.7268769741 \\
      0.42 &0.07861837136 & 0.7270777797 \\
      0.47 &0.07950720626 & 0.7287969021
    \end{tabular}
    \caption{Evolution of $\Re(t)$ at $t_{max}$ for various $\alpha$ in the strip region for AdS-Gauss-Bonnet bulk.} 
  \end{center}
\end{table}
\section{Thermalization Regimes after Quench}
In this section we are about to study the entanglement growth
discussed in the previous section in more details. We can see
there are some similarities and differences between time evolution
of entanglement entropy for small and large subsystems before
saturation. As we can see for large systems in [19,20], three
distinct regimes have been studied separately:\\

 \textit{1)
pre-local equilibration regime} which happens at initial times of
quench for which $t\ll t_{loc}\sim z_h$, in this regime
entanglement entropy behaves like a quadratic function of time,\\

\textit{2) post-local equilibration regime} for $t\geqslant
t_{loc}$, in this time interval the growth function takes a linear
shape, and\\

 \textit{3) approach to saturation} which studies this
behavior close to the saturation time and also phase transition of
the equilibrium of entanglement entropy.\\

 As mentioned above in
the large subsystems local thermal equilibrium time, ($t_{loc}$),
is an important timescale and has been used to study intermediate
regime of the evolution. The problem arises in small subsystems,
because these systems saturate long before
 it equilibrated thermally, in fact since $z_t\ll z_h$ so $t_{sat}\ll t_{loc}$, so we don't have any definite timescale like in the large subregion
 case. In addition there are some differences as well, in the universality during various time intervals. Similar to large subsystem case, we will
  study the entanglement growth in small region case ($z_t\ll z_h$), during three distinct time intervals as follows:\\

$\bullet$ \textbf{Initial quadratic growth:} Since the only
time-scale in this approximation is the saturation time, then for
initial time we have $t\ll t_{sat}$ corresponds to $t\ll z_t$. We
can apply this limitation in our results for our gravity model.
From (4.25) we obtain:
\begin{equation}
\Delta S(t\ll t_{sat})=\frac{(2-\alpha)^{3/2}A_{\Sigma}}{32\sqrt{2}G_N^{(5)}z_h^4}~t^2+...
\end{equation}
for $\alpha$ defined in (4.9). This time evolution function (as it
is expected) is quadratic for initial time and independent of the
size region, it also must presents an universal behavior.
Therefore for any other shapes of entangled region, the same
result would be
 obtained in this regime.\\

$\bullet$ \textbf{Quasi-linear growth:} In the large subsystems after the local equilibrium point, time evolution of the regime would be
 linearly which is the opposite of the behavior in our case. By attention to the equation (4.27), entanglement growth is not obviously
 linear so tsunami velocity (which is defined in large subsystems) is useless here. In the other side, because instantaneous rate $\Re(t)$
 depends on the size and shape of the entangled region so the evolution is not universal. This is obvious from (4.29) that we can see $z_t$
 in it which is related to the size of region. By attention to this fact that $\Re_{max}=v^{max}_{E}$ as it described in introduction, we can
  produce an equation very similar to the linear regime for large subsystems:
\begin{equation}
\Re_{max}=v^{max}_{E}=\frac{1}{s_{eq}A_{\Sigma}}\frac{d(\Delta S_A(t))}{dt}\bigg|_{t=t_{max}},
\end{equation}
which is just valid for $t_{max}$. At this time instantaneous value of the rate takes its maximum amount which could be greater than
 1. By a mathematical approach we get to,
\begin{equation}
\Delta S_A(t)-\Delta S_A(t_{max})=v^{max}_{E}s_{eq}A_{\Sigma}(t-t_{max})+\mathcal{O}(t-t_{max})^3.
\end{equation}
Since there is not any quadratic term in the above equation,
therefore the time behavior of entanglement entropy will be linear
at $t_{max}$. In this quasi-linear regime we can evaluate
$v_E^{max}$ at the moment $t_{max}$ for our gravity model.
Extremizing the instantaneous rate (4.29) leads to the following
real root which defines for acceptable values of $\alpha$:
\begin{equation}
t_{max}=z_tf(\alpha),
\end{equation}
where,
\begin{equation}
f(\alpha)=\frac{\sqrt{2\alpha\big(5\alpha-2+\sqrt{9\alpha^2-4\alpha+4}\big)}}{4\alpha}.
\end{equation}
By putting this value in (4.29) we can evaluate the maximum rate of the growth as follows:
\begin{equation}
v_E^{max}=\Re(t)\bigg|_{t_{max}}=\frac{3f(\alpha)\sqrt{1-f^2(\alpha)}}{4\big(1+\frac{2\alpha}{5(1-\alpha)}\big)}
\bigg(1+\frac{\alpha}{1-\alpha}f^2(\alpha)\bigg)\sqrt{2(2-\alpha)},
\end{equation}
which implies that for a definite size of region, $v_E^{max}$ just depends on our gravity model. Of course it must be noted
that $v_E^{max}$ by attention to $\Re(t)$ depends on the shape and size of entangled region which means it would not be universal
and also not bounded by $v_E\leqslant1$, generally. By attention to table 1 and figure (1.b) we can see that this value never exceeds
 the speed of light. \\

$\bullet$ \textbf{Just before saturation:} By attention to [19,20] the evolution of entanglement entropy just before saturation time
in large subsystems depends on the shape of region, the final state and dimension of spacetime. By approaching to $t_{sat}$ the system
bears a phase transition to equilibrated state at which the evolution of entanglement ceased. The kind of this phase transition depends
on the shape of region as in the strip case for $d\geqslant3$ this transition occurs suddenly, or in mathematical language time derivative
of $\Delta S(t)$ would be discontinuous at saturation time which corresponds to a first order phase transition. On the contrary, for any
 dimension the derivative of $\Delta S(t)$ at $t_{sat}$ is continuous for the ball region which corresponds to a second order phase transition.
 In this kind of phase transition the behavior of the system can be characterized by a nontrivial scaling exponent $\gamma$ for which,
\begin{equation}
\Delta S_A(t)-\Delta S_A(t_{sat})\varpropto(t_{sat}-t)^\gamma,~~~~~\gamma=\frac{d+1}{2},
\end{equation}
where $(t_{sat}-t)\ll\ell_{eq}$. This evolution is valid for any dimensions even for BTZ black hole with $d=2$, but in $d=3$ the evolution takes
the form $\Delta S_A(t)-\Delta S_A(t_{sat})\varpropto(t_{sat}-t)^2\log{(t_{sat}-t)}$ for which the logarithmic term makes it different from
 the mean-field behavior with $\gamma=2$ in standard thermodynamic transitions.\\

The situation in small subsystems is not necessarily same as large ones. For various regions time derivative is continuous and then phase
 transition is of second order. By expanding $\Delta S(t)$ around the saturation time, critical exponent would be different in various entangled
  regions. In our gravity model the critical exponent would be achieved as $\gamma=3/2$
  .\\

Discontinuity in entanglement entropy for large subsystems around the saturation time comes from the multi-valuedness of turning point at this
 time which causes to a swallow-tail behavior of entanglement entropy. But in small regions the situation is different and the leading order
 contributions comes from the pure AdS and causes to a unique value of $z_t$ at saturation time which makes phase transition continuous in contrary
 with large subsystems. Of course it must be noted that by considering higher orders of entropy and appearing time dependent parts in addition to pure
 AdS, we expect that this multi-valuedness will appear.\\

\section{Conclusion}
$\bullet{\textbf{~Results:}}$  In the present work we tried to
study different regimes of entanglement entropy evolution for a
non -Einstein-like gravity model. We are interested in analytical
answers instead of numerical solution, so we use an approximation
approach which bounds the case to small subsystems on CFT side.
Thermalization regimes by this approximation would be computed at
the leading order terms. Tsunami picture breaks down here because
of small region approximation in which the saturation happened
long
 before the thermal excitations could be effective, so the instantaneous rate is not bounded by the speed of light. Studying the details
  of thermalization for entanglement entropy in this model shows that the evolution is universal for initial times after quench which is
   the same as the large subsystems case. But this similarity disappears at intermediate times while in large subsystems the evolution
   is linear after local thermal equilibrium time but in small region the situation is different. The rate of entanglement growth in
   small subsystems has a maximum value for a specified time in the intermediate regime which can be varied by changing the size of the
   region. In the model which we used here one can infer by increasing $\lambda_{GB}$ the maximum value of the rate or $v_E^{max}$
   reaches to decreasing behavior and then an increasing at late times whilst the time at which $v_E^{max}$ occurs will be delayed more and more,
   it could be seen also that $v_E^{max}$ never exceeds the speed of light.\\

Another important result is that the entanglement entropy function approaching to the equilibration point at saturation
 time leads to a continuous derivative. So phase transitions would be of second order and we can characterize it with scaling exponent.
 It must be noted that this scaling exponent which is a scale factor in continuous phase transitions could be varied in various gravity models
 and for different shapes of entangled regions.\\

$\bullet{\textbf{~Purposes and the future works:}}$  An analytical solution of entanglement growth makes it possible studying the various
regimes of thermalization in more details from the point the quench turns on until saturation time. We can investigate similarities and
 differences of this evolution among various gravity models and with different shape and size of entangled subregion on the boundary [31],
  or we can also study the effects of characteristic parameters of our gravity model. It would be useful studying the behavior of system
  approaching to saturation time and the kind of phase transition with an explicit function of its evolution. Actually some numerical
  approaches have done on small and large subsystems for various gravity models in [25,32,33] and phase transition structure have studied
  in [34,35] which our results could be compared with them.\\

As a future work we can study the case in the context of anti-de
Sitter/condensed matter theory (AdS/CMT) correspondence in which
the dual gravity model has different symmetries such as theories
with Lifshitz scaling or hyperscaling violation. Studying phase
structure of black holes by employing non-local observables like
entanglement entropy, Wilson loop or two point functions in the
static case as a probe could be another interesting area. It must
be noted that the main motivation for this study comes from this
question whether these observables have a phase structure like
thermal entropy of black holes in the $T-S$ plane or not. Also we
can generalize this topic to $d$-dimension and study the effects
of the dimension on the behavior of non-local observables and
thermalization regimes. We can also generalize the gravity model
to general higher derivative Lovelock theories. There are also
other observables which can be more applicable such as: n-partite
information like mutual information and studying monogamy and
other properties of entanglement, casual holographic information
and holographic complexity which the latter is in progress in our
next work. We can also consider an arbitrary profile for the mass
or charge in quench process and write down the linear response for
entanglement entropy of small subsystems and close to the vacuum
[36].

\appendix

\section{Gibbons-Hawking-York term}
The induced metric on the surface $\Sigma$ from (5.2) is given by
\begin{equation}
ds^2=\gamma_{ab}dx^adx^b=\mathcal{N}^2dz^2+h_{ij}dx^idx^j=\mathcal{N}^2dz^2+\frac{L^2}{z^2}(dx_2^2+dx_3^2),
\end{equation}
in which $h_{ij}$ is the components of the induced metric on the boundary of entangled surface $\Sigma$ (or first fundamental form) and
\begin{equation}
\mathcal{N}^2=\frac{L^2\mathcal{G}}{z^2f_0}.
\end{equation}
Actually the induced metric on the boundary $\partial\Sigma$ is related to the induced metric on the surface $\Sigma$ through
\begin{equation}
h_{ab}=\gamma_{ab}+n_an_b;~~~n_1=n_z=-\mathcal{N},~~~n_i=0,
\end{equation}
where $n^1=n^z=-1/\mathcal{N}$ is unit normal vector on $z$ direction and obviously other components are zero:
\begin{equation}
n^a=\bigg(-\frac{z\sqrt{f_0}}{L\sqrt{\mathcal{G}}},0,0\bigg).
\end{equation}
By these definitions the determinant of the induced metric at the
boundary is $\sqrt{h}=\frac{L^2}{z^2}$ and by attention to [37]
the mean curvature (or the trace of extrinsic curvature) is given
by:
\begin{equation}
K=h^{ij}K_{ij},
\end{equation}
in which $K_{ij}=\nabla_in_j$ is the components of extrinsic
curvature (or second fundamental form) and by attention to zero
value of the \textit{shift} $N^{\alpha}$ (for details see again
[37] chapter 12) obtained as below:
\begin{equation}
K_{ij}=-\frac{1}{2}n^a\partial_ah_{ij}.
\end{equation}
Finally the above relations lead us to the trace of extrinsic curvature as:
\begin{equation}
K=-\frac{2\sqrt{f_0}}{L\sqrt{\mathcal{G}}}.
\end{equation}

  \vskip .5cm
 \noindent
  {\bf References}
\begin{description}
\item[1.] J. Maldacena, "The Large N Limit of Superconformal Field Theories and Supergravity," Adv. Theor. Math. Phys. 2 (1998) 231, hep-th/9711200.
\item[2.] E. Witten, "Anti De Sitter Space And Holography," Adv. Theor. Math. Phys. 2 (1998) 253, hep-th/9802150.
\item[3.] G. 't Hooft, "Dimensional Reduction in Quantum Gravity," gr-qc/9310026.
\item[4.] S. A. Hartnoll, "Lectures on holographic methods for condensed matter physics," Class. Quant.
Grav. 26, 224002 (2009),hep-th/0903.3246.
\item[5.] S. Ryu and T. Takayanagi, "Holographic derivation of entanglement entropy from AdS/CFT," Phys. Rev. Lett. 96, 181602 (2006),
hep-th/0603001.
\item[6.] V. E. Hubeny, M. Rangamani, and T. Takayanagi, A Covariant holographic entanglement entropy proposal, JHEP 0707 (2007) 06;
hep-th/0705.0016.
\item[7.] B. Freivogel, R. A. Jefferson, L. Kabir, B. Mosk, and I.-S. Yang, Casting Shadows on Holographic Reconstruction, Phys. Rev. D 91, 086013 (2015),
hep-th/1412.5175.
\item[8.] A. Allais and E. Tonni, Holographic evolution of the mutual information, JHEP 01
(2012) 102,hep-th/1110.1607.
\item[9.] V. E. Hubeny and M. Rangamani, Causal Holographic Information, JHEP 1206 (2012)114, hep-th/1204.1698.
\item[10.] M. Headrick and T. Takayanagi, "A holographic proof of the strong subadditivity of entanglement entropy," Phys. Rev. D 76, 106013 (2007),
hep-th/0704.3719.
\item[11.] L.-Y. Hung, R. C. Myers, and M. Smolkin, On Holographic Entanglement Entropy and Higher Curvature Gravity, JHEP 1104 (2011) 025,
hep-th/1101.5813.
\item[12.] J. de Boer, M. Kulaxizi, and A. Parnachev, Holographic Entanglement Entropy in Lovelock Gravities, JHEP 1107 (2011) 109, hep-th/1101.5781.
\item[13.] X. Dong, Holographic Entanglement Entropy for General Higher Derivative Gravity, JHEP 1401 (2014) 044, hep-th/1310.5713.
\item[14.] J. Camps, Generalized entropy and higher derivative Gravity, JHEP 1403 (2014) 070, hep-th/1310.6659.
\item[15.] U. H. Danielsson, E. Keski-Vakkuri and M. Kruczenski, "Spherically collapsing matter
in AdS, holography, and shellons," Nucl. Phys. B 563, 279 (1999),
hep-th/9905227.
\item[16.] U. H. Danielsson, E. Keski-Vakkuri and M. Kruczenski, "Black hole formation in AdS and thermalization on the boundary," JHEP 0002, 039 (2000),
hep-th/9912209.
\item[17.] S. B. Giddings and A. Nudelman, "Gravitational collapse and its boundary description in AdS," JHEP 0202, 003 (2002),
hep-th/0112099.
\item[18.] P. Calabrese and J. L. Cardy, "Evolution of entanglement entropy in one-dimensional systems," J. Stat. Mech. 0504, P04010 (2005),
cond-mat/0503393.
\item[19.] H. Liu and S. J. Suh, Entanglement growth during thermalization in holographic systems,",Phys. Rev. D 89, 066012 (2014),  hep-th/1311.1200.
\item[20.] H. Liu and S. J. Suh, Entanglement Tsunami: Universal Scaling in Holographic Thermalization,"Phys. Rev. Lett. 112, 011601 (2014), hep-th/1305.7244.
\item[21.] H. Casini, H. Liu and M. Mezei, "Spread of entanglement and causality," JHEP 1607,077 (2016), hep-th/1509.05044.
\item[22.] T. Hartman and N. Afkhami-Jeddi, "Speed Limits for Entanglement," hep-th/1512.02695.
\item[23.] X. Zeng and W. Liu, "Holographic thermalization in Gauss-Bonnet gravity," Phys. Lett. B726, 481 (2013),hep-th/1305.4841.
\item[24.] R. G. Cai, Gauss-Bonnet Black Holes in AdS Spaces, Phys. Rev. D 65, 084014 (2002),hep-th/0109133.
\item[25.]  E. Caceres, M. Sanchez and J. Virrueta, "Holographic Entanglement Entropy in Time Dependent Gauss-Bonnet Gravity,"
hep-th/1512.05666.
\item[26.] H. Ghaffarnejad, `Classical and
Quantum Reissner-Nordstr\"{o}m Black Hole Thermodynamics and first
order Phase Transition`,Astrophys. and Space Sci. 361 , 1 (2016),
physics.gen-ph/1308.1323.
 \item[27.] S. G.
Ghosh, S. D. Maharaj and U. Papnoi, `Radiating Kerr-Newman black
hole in $f(R)$ gravity`,Eur. Phys. J. C (2013) 73: 2473,
gr-qc/1208.3028.
\item[28.] X. X. Zeng and L. F. Li, `Van
der Waals phase transition in the framework of holography`, Phys.
Lett. B764, 100-108 (2017),hep-th/1512.08855.
\item[29.] S. Kundu and J. F. Pedraza, Spread of entanglement for small subsystems in holographic CFTs," Phys. Rev. D 95, 086008 (2017),
hep-th/1602.05934.
\item[30.] G. Gibbons and S. Hawking, "Action Integrals and Partition Functions in Quantum Gravity," Phys.Rev. D15 (1977) 2752–2756.
\item[31.]  T. Albash and C. V. Johnson, "Evolution of holographic entanglement entropy after thermal and electromagnetic quenches,â€
 New Journal of Physics 13 no. 4, (2011) 045017.
\item[32.] V. Balasubramanian et al., "Thermalization of Strongly Coupled Field Theories," Phys.Rev. Lett. 106, 191601 (2011), hep-th/1012.4753.
\item[33.] V. Balasubramanian et al., "Holographic Thermalization," Phys. Rev. D 84, 026010(2011), hep-th/1103.2683.
\item[34.] S. He, L.-F. Li and X.-X. Zeng, "Holographic Van der Waals-like phase transition in the Gauss-Bonnet gravity," Nucl. Phys. B915 (2017)
 243–261,hep-th/1608.04208.
\item[35.]  X.-X. Zeng and L.-F. Li, "Van der Waals phase transition in the framework of holography," Phys. Lett.B764(2017) 100-108,hep-th/1512.08855.
\item[36.] S. F. Lokhande, G. W. J. Oling, and J. F. Pedraza, "Linear response of entanglement entropy from holography," JHEP 10 (2017) 104
,hep-th/1705.10324.
\item[37.] T. Padmanabhan, "Gravitation: Foundations and frontiers," Cambridge, UK: Cambridge University Press. (2010).
\end{description}

\begin{figure}[ht]
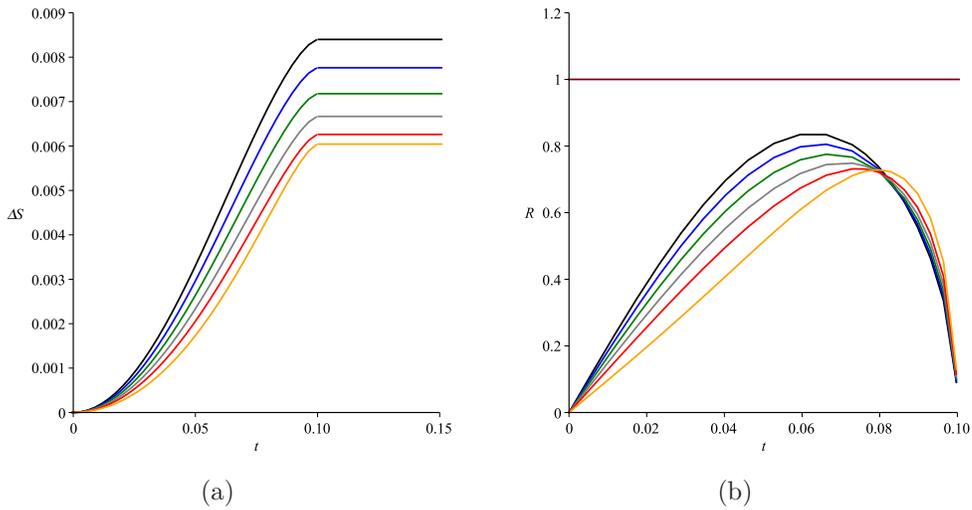

\centering
\subfigure[{}]{\label{1}
\includegraphics[width=.45\textwidth]{1.eps}}
\hspace{3mm}
\subfigure[{}]{\label{1}
\includegraphics[width=.45\textwidth]{2.eps}}
\hspace{3mm}
\caption{$(a)$ Entanglement entropy evolution for a small subregion in Gauss-Bonnet gravity, where we set $z_t/z_h=0.1$ due to small subregions and also $z_h=1$ and $A_{\Sigma}/16G_N^{(5)}=1$ for simplicity. In these diagrams we put $\alpha=-0.6,-0.4,-0.2,0,0.2,0.47$ which indicates by black, blue, green, gray, red and orange lines. The case with $\alpha=0$ represents when higher order Gauss-Bonnet term vanishes. In diagram $(b)$ we plotted instantaneous rate of this evolution for which the values of gravity model parameter is indicated by the same colors. The horizontal line is the constant speed of light. }
\label{l}
\end{figure}

\end{document}